# Highly Efficient Second/Third Harmonic Generation in van der Waals Layered Material AgScP$_2$S$_6$ with Anisotropic Polarization and Temperature Dependence

*Mohamed Yaseen Noor\*, Ryan Siebenaller, Wenhao Liu, Zixin Zhai, Conrad Kuz, Simin Zhang, Mousumi Upadhyay Kahaly, Gergely Nagy, Aamir Mushtaq, Rahul Rao, Emmanuel Rowe, Benjamin S. Conner, Bing Lv, Michael A. Susner\*, Enam Chowdhury\**

Mohamed Yaseen Noor, Ryan Siebenaller, Conrad Kuz, Simin Zhang, Aamir Mushtaq and Enam Chowdhury

Materials Science and Engineering, The Ohio State University, Columbus, OH 43202, USA

Wenhao Liu, Zixin Zhai, and Bing Lv

Department of Physics, the University of Texas at Dallas, Richardson, TX 75080, USA

Mousumi Upadhyay Kahaly, and Gergely Nagy

ELI ALPS, ELI-HU Non-Profit Ltd., Wolfgang Sandner utca 3., Szeged 6728, Hungary

Rahul Rao, Emmanuel Rowe, Benjamin S. Conner, and Michael A. Susner

Materials and Manufacturing Directorate, Airforce Research Laboratory, Wright-Patterson AFB, OH 45433, USA

Benjamin S. Conner

Azimuth Corporation, Core4ce LLC, Fairborn, Ohio 45342, USA

E-mail: noor.48@osu.edu, chowdhury.24@osu.edu, michael.susner.1@us.af.mil

**Funding:** United States Air Force Office of Scientific Research (AFOSR) LRIR23RXCOR003, LRIR26RXCOR010, and AOARD-NSTC Grant Number F4GGA21207H002. United States AFOSR Grant No. FA9550-19-1-0037, National Science Foundation- DMREF-2324033, and Office of Naval Research Grant No. N00014-23-1-2020.

**Abstract:**

Single-crystal X-ray diffraction and nonlinear optical measurements, especially second- and third-harmonic generation (SHG/THG) are comprehensively investigated for the van der Waals layered material AgScP$_2$S$_6$ with a non-centrosymmetric P31c (159) space group. Linear optical constants are extracted using spectroscopic ellipsometry and applied in fitting the harmonic generation behavior. Polarization-resolved SHG and THG measurements exhibit pronounced anisotropy, with emission patterns well-described by theoretical models derived from the $\chi^2$ and $\chi^3$ tensor elements. The material demonstrates exceptionally high nonlinear susceptibilities, with $\chi^2 \approx 10^{-8}$ m/V and $\chi^3 \approx 10^{-17}$ m$^2$/V$^2$ which is a few orders of magnitude greater than comparable 2D materials reported in the literature. Temperature-dependent SHG and THG measurements from 300 K to 25 K reveal exponential decay in harmonic signal intensities, attributed to reduced carrier mobility, with no evidence of structural phase transitions, consistent with results from single crystal diffraction and heat capacity measurements. Polarization-resolved SHG and THG measurements also reveal distinct orientation and ellipticity trends, highlighting the anisotropic nonlinear tensor contributions and contrasting polarization selection rules in the material. These results establish $AgScP_2S_6$ as a high-performance, thermally stable, and highly anisotropic nonlinear candidate material suitable for compact photonic applications such as ultrafast optical modulators, polarization-sensitive detectors, and wavelength-tunable light sources.

1. Introduction

The development of new compact nonlinear photonic devices is driven by materials that exhibit strong light–matter interactions combined with broad optical absorption characteristics and enhanced efficiency in nonlinear conversion processes [1]. Van der Waals (vdW) materials stand out as exemplars of this trend, due to their reduced dimensionality and high surface-to-volume ratio leading to enhanced non-linear effects and offering a versatile platform for such devices [1–3]. These devices are finding applications across various fields, including ultrafast optical modulators, wavelength-sensitive multiplexers, rapid photonic circuits, and lasers utilizing mode-locking or Q-switching techniques [4,5]. Moreover, in-plane anisotropy in two-dimensional (2D) materials, arising from their reduced crystal symmetry, offers an additional avenue for manipulating their electrical, optical, thermal, or mechanical characteristics. 2D materials such as black phosphorus (BP), GeSe, ReS$_2$, InSe, and SnSe$_2$ have garnered increasing interest due to their pronounced in-plane anisotropy [6–11], while isotropic layered materials like MoS$_2$ serve as useful reference systems. Combining this anisotropy with other inherent properties derived from the layered structure of 2D materials presents novel opportunities for developing photonic devices tailored for specific applications [12,13].

The in-plane anisotropy in two-dimensional (2D) materials originates from their reduced crystallographic symmetry, which in turn gives rise to direction-dependent variations in orbital overlap, bond lengths, and local bonding environments. As a result, properties such as carrier mobility, optical absorption, thermal conductivity, and mechanical elasticity become intrinsically direction dependent. Black phosphorous and other novel 2D materials have been extensively studied for their anisotropic properties, particularly in nonlinear optical processes such as high harmonic generation (HHG), saturable absorption (SA), and four-wave mixing (FWM) etc., [14,15]. However, their practical utility is hindered by their chemical instability in ambient conditions and low non-linear optical (NLO) responses [16]. It is in this milieu that $AgScP_2S_6$ emerges as a promising and notably top 2D candidate for nonlinear applications. $AgScP_2S_6$ is a unique 2D material: its van der Waals layered structure, combined with a complex lattice featuring multiple atomic species, variable bond lengths, and nontrivial angular distortions, generates pronounced in-plane anisotropy. Recent studies have shown that $AgScP_2S_6$ remains chemically stable even at few-layer thicknesses, outperforming black phosphorus, and similar anisotropic crystals whose degradation in air limits device integration [17,18]. This combination of intrinsic anisotropy, large two-photon absorption coefficient, and ambient stability down to thin-flake limits positions $AgScP_2S_6$ as an attractive platform for nonlinear photonics.

However, to fully harness the advantages of nonlinear applications, a detailed understanding of its behavior under varying polarization and temperature conditions is essential. High harmonic generation (particularly second (SHG) and third harmonic generation (THG)) has emerged as a powerful tool in nonlinear optics, with applications spanning deep-tissue bioimaging, frequency conversion, and ultrafast optical diagnostics [19,20]. These processes involve the interaction of two or three photons with a material to produce a single photon at half or one-third the wavelength, enabling high-resolution imaging with minimal photodamage. In atomically thin crystals, the vanishing propagation length removes traditional phase-matching constraints; the nonlinear polarization generated within the material remains coherent across the entire thickness, allowing efficient harmonic emission even without engineered dispersion compensation. This intrinsic "phase-matching freedom" is a defining advantage of 2D systems and contributes to their large effective $\chi^2$ and $\chi^3$ responses [21]. Additionally, the generation of chiral light through harmonic generation has significant implications for the study of molecular chirality [22], a property of asymmetry in molecules that plays a crucial role in various biological processes and the development of pharmaceuticals [20,23].

In this study, we investigate the nonlinear optical properties of $AgScP_2S_6$. First, we characterize the linear optical constants of the material using spectroscopic ellipsometry which thus provides essential information for our nonlinear measurements. We then employ rotational anisotropy harmonic generation (RAHG) measurements to determine the crystallographic orientation of bulk $AgScP_2S_6$, by rotating the incident field's polarization while keeping the sample fixed and analyzing the polarization-resolved SHG and THG signals. We fit the experimental data into a generalized theoretical model that accounts for polarization dependent non-linear responses. We then quantify the SHG and THG efficiencies and extract the second- and third-order nonlinear susceptibilities, revealing values that exceed those reported for many 2D materials. Our temperature dependent measurements of the harmonic signal provide insights into crystal symmetry, phase transitions, and electron-phonon interactions. Additionally, the polarization states of the harmonic signals are analyzed; signal optimization is demonstrated by tuning the fundamental beam's polarization from linear to circular. By examining the effects of polarization, temperature, and chirality dependencies, this work aims to elucidate the mechanisms governing harmonic generation in AgScP$_2$S$_6$, thereby contributing to the development of next generation nonlinear photonic technologies.

## 2. Results

### 2.1 AgScP$_2$S$_6$ Crystal Characterization

AgScP$_2$S$_6$ is a class of exotic 2D materials known as layered quaternary metal thiophosphates. Structurally, a single lamella of layered quaternary metal thio/selenophosphate consists of closely packed sulfur/selenium atoms that surround octahedral sites in which metal cations (2/3 of these sites) and P-P pairs (the remaining 1/3 of these sites) are embedded. These layers are stacked together vertically via weak van der Waals interaction to form the bulk crystal. The only known previous report [24] of AgScP$_2$S$_6$ in literature identifies the crystal space group of *P-31c*, a centrosymmetric trigonal structure. SHG emission from bulk crystals pointed us to question this structural assignment, which we have validated through single crystal XRD and detailed SHG studies. Our refinement identifies that, instead, AgScP$_2$S$_6$ falls in the *P31c* space group (#159) and thus loses its previously reported inversion center. Our refinements accounted for the presence of racemic twinning which helped to delineate the new space group from the previously reported literature. Both room temperature and 100K structural refinements can be seen in Tables 1 and 2, respectively.

**Table 1.** Room Temperature and 100K structural refinements of AgScP$_2$S$_6$ from X-ray single crystal diffractions.

| Lattice constants | Site | Atomic coordinates | | | $U_{ani}$ (Å$^2$) | Occ. f. |
|---|---|---|---|---|---|---|
| | | x | y | z | | |
| a = b = 6.1749(2) Å | Ag1 | ⅔ | ⅓ | 0.4685(9) | 0.0341(15) | 0.29(2) |
| c = 12.9056(4) Å | Ag1' | ⅔ | ⅓ | 0.5034(6) | 0.0341(15) | 0.71(2) |
| α = β = 90° | Sc1 | 1 | 1 | 0.5002(3) | 0.0117(4) | 1 |
| γ = 120° | P1 | ⅓ | ⅔ | 0.5854(3) | 0.0125(9) | 1 |
| *P31c* (No. 159) | P2 | ⅓ | ⅔ | 0.4147(4) | 0.0083(7) | 1 |
| | S1 | 0.6620(4) | 0.6950(5) | 0.62791(13) | 0.0128(5) | 1 |
| | S2 | 0.3053(6) | 0.3385(4) | 0.36977(14) | 0.0162(6) | 1 |
| **Reliability Factors** | | $R_1$ | $wR_2$ | $R_{int}$ | GooF | |
| | | 0.0355 | 0.0980 | 0.0376 | 1.366 | |

Notably the Ag position in this system is split into two sites with 71/29 occupancy, slightly above and below the midline of the lamella, near the center of the octahedral site. As other MTP compounds are known to show complex temperature dependences on monovalent cation ordering that in turn govern the ferroic properties of these compounds (e.g. [25–29]), we have explored structural refinements and characterization across high and low temperatures. We find no significant difference in structural refinement as low as 100 K, and no evidence of structural shifts at low temperatures is seen in our other characterization methods such as differential scanning calorimetry (DSC) and specific heat capacity measurements.

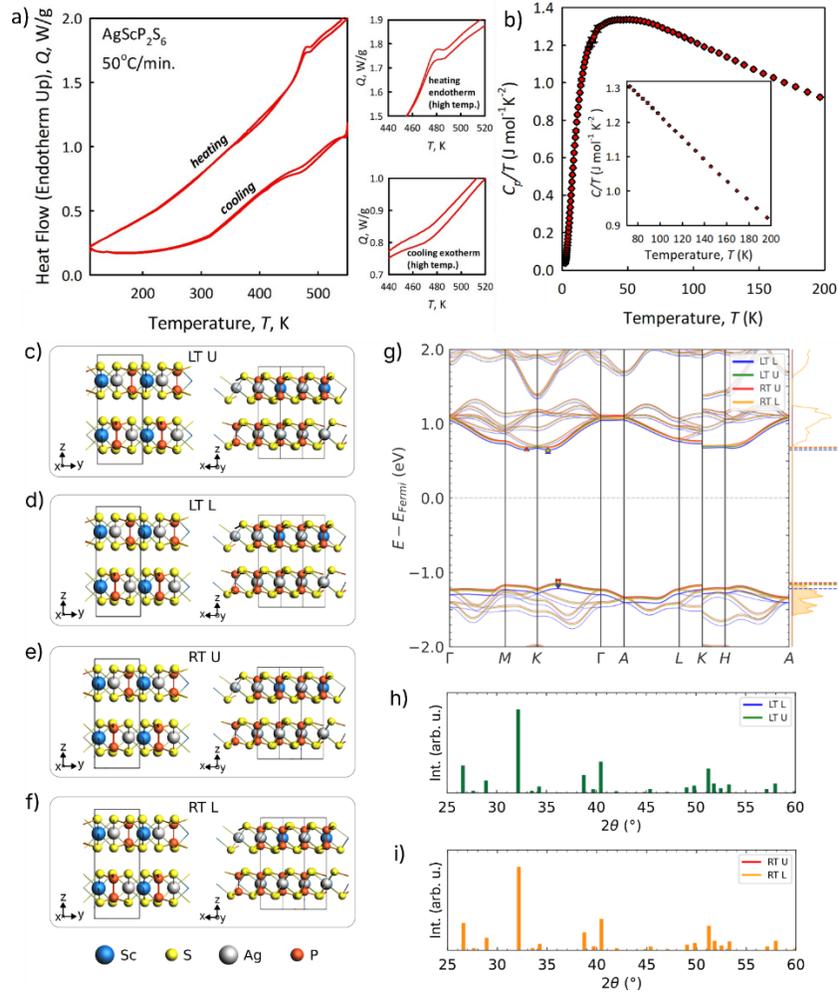

**Figure 1:** DSC endotherm **(a)** showing no notable low temperature phase transition on heating or cooling curves. Inset show a likely high temperature phase transition on heating and cooling near 480 K, outside the range of our other testing methods. Heat capacity measurements **(b)** taken on a single crystal again showing no clear phase transitions. Inset on the temperature range in which we note a slight shift in SHG intensity. **(c-f)** Optimized 2×2×1 supercells AgScP$_2$S$_6$ starting from different initial geometries for low-temperature (LT) and room-temperature (RT) phases. However, after energy and force minimization, they reach almost identical structures. (g) Corresponding electronic band structure of the phases, further indicating very similar ground state electronic properties. (h-i) Simulated XRD patterns for optimized LT and RT phases, indicating structural phase stability and thermal stability of the crystal.

DSC analysis of $AgScP_2S_6$ reveals a distinct reversible endothermic transition near 480 K, observed in both heating and cooling curves at a scan rate of 50 °C/min. Insets highlight the endothermic peak on heating and the corresponding exothermic dip on cooling, suggesting a first-order transition. This thermal anomaly is further confirmed by variable-temperature powder XRD measurements off the (001) face (see Fig. S2), which show a sharp discontinuity in layer spacing at ~480 K defined as the sum of the lamella thickness and van der Waals gap followed by a reduced slope beyond the transition. Notably, this transition occurs well below decomposition temperatures typical of MTPs (>1000 K), yet higher than most known phase transitions in such materials, with the exception of the cation melting transition in $In_{4/3}P_2S_6$ (945 K) and the CIPS-IPS heterostructure (520 K) [25]. Specific heat measurements ($C_p/T$) show no thermodynamic anomalies between 1.9 K and 400 K, indicating the absence of low-temperature phase transitions.

The atomic-scale origin of the high-temperature transition was further probed via density functional theory (DFT) calculations. Experimental structural refinement reveals that Ag atoms in $AgScP_2S_6$ occupy two distinct positions in both the low-temperature (LT) and room-temperature (RT) phases: ~70% reside in an "upper" (U) site, while ~30% occupy a slightly lower "L" site within the $S_6$ coordination cage. To assess the energetic stability of these configurations, we performed DFT structure optimizations using the AMS package (DZP basis, GGA-PBE functional, 3×3×3 k-point sampling), assuming all Ag atoms were initially in either U or L sites. In all cases, the L-site configurations relaxed to the U-site geometry, suggesting that the L sites are metastable or unstable in isolation. To more accurately reflect the experimentally observed mixed occupancy, we constructed a 2×2×1 supercell (80 atoms/unit cell) containing eight Ag atoms, with two symmetrically positioned in L sites (25% L: 75% U). These mixed-occupancy supercells retained structural stability and allowed for further electronic structure analysis. The resulting relaxed geometries and their corresponding band structures for both LT and RT phases are shown in Fig. 1(c), providing insight into the subtle electronic consequences of Ag site disorder and the nature of the phase transition.

Bulk $AgScP_2S_6$ crystals were grown using chemical vapor transport method, a more detailed description of growing methods of these vdW layered materials has been published in our previous works [17,30]. The crystal structure and the optical image of the bulk $AgScP_2S_6$ used for this study are shown in Figure S1.

**Table 2.** 100 K refinements of AgScP$_2$S$_6$.

| Lattice constants | Site | Atomic coordinates | | | U$_{ani}$ (Å$^2$) | Occ. f. |
|---|---|---|---|---|---|---|
| | | x | y | z | | |
| a = b = 6.1613(3) Å | Ag1 | ⅔ | ⅓ | 0.4738(8) | 0.0085(12) | 0.30(2) |
| c = 12.8701(9) Å | Ag1' | ⅔ | ⅓ | 0.5022(5) | 0.0085(12) | 0.70(2) |
| α = β = 90° | Sc1 | 1 | 1 | 0.4998(4) | 0.0049(4) | 1 |
| γ = 120° | P1 | ⅓ | ⅔ | 0.5854(3) | 0.0074(9) | 1 |
| P31c | P2 | ⅓ | ⅔ | 0.4147(4) | 0.0029(4) | 1 |
| | S1 | 0.6623(6) | 0.6937(8) | 0.6271(2) | 0.0060(8) | 1 |
| | S2 | 0.3057(8) | 0.3382(6) | 0.3680(2) | 0.0162(6) | 1 |
| Reliability Factors | | $R_1$ | $wR_2$ | $R_{int}$ | GooF | |
| | | 0.0361 | 0.1022 | 0.0371 | 1.537 | |

Sample composition was verified via energy-dispersive spectroscopy (EDS), and structural characterization was conducted using temperature-dependent powder and single-crystal X-ray diffraction (XRD). Additional thermal analysis included differential scanning calorimetry (DSC) and specific heat measurements to investigate potential phase transitions. Temperature-dependent Raman spectroscopy was also performed along the stacking face of AgScP$_2$S$_6$ flakes to probe vibrational modes.

### 2.2 Non-Linear Measurements

Anisotropic second- and third-harmonic generation in AgScP$_2$S$_6$ was produced using a 1030 nm, 70 fs pulsed laser at 2 kHz with ~0.68 TW/cm$^2$ peak intensity, and the emitted harmonics were collected via short-pass filtering and routed to a visible spectrometer. Figure 2 (a) & (b) shows measured second harmonic (SHG) and third harmonic (THG) in linear and log scale respectively. To ensure the SHG and THG emission are being collected, the emission intensity is recorded as a function of input pump power and fitted with a linear function in a log$_{10}$-log$_{10}$ plot as shown in Figure 2 (c). The measured SHG and THG intensity values are indicated with green and blue circles respectively, while the linear fit is shown in dotted lines. The fit slopes of 1.89 and 3.28 indicate that the SHG is a second order process whereas the THG emission is a third order process. Notably, the intensity maximum exhibited a peak at 515 nm and 343 nm, precisely one-half and one-third of the incident pump wavelength. The conversion efficiency of SHG/THG from a ~9 μm thick AgScP$_2$S$_6$ crystal is few orders of magnitude higher than that of other reported vdW layered materials like MoSe$_2$, WS$_2$ when normalized to thickness [31–34]. The SHG conversion efficiency measured here for the average pump power of 2 mW, which corresponds to a peak intensity of 0.68 TW/cm$^{-2}$ is ~1.55x10$^{-2}$, whereas the THG conversion efficiency for same input peak intensity is ~5.25x10$^{-5}$.

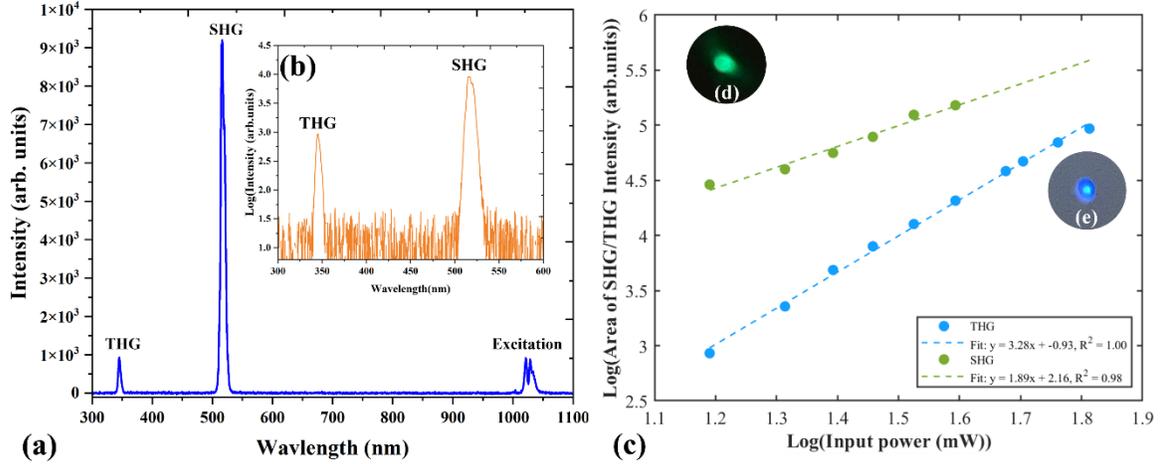

**Figure 2: (a)** Measured SHG and THG intensity of AgScP$_2$S$_6$ crystal when excited at 1030nm with an average power of 50 mW or peak intensity of 110 TW/cm$^2$ at focus. **(b)** SHG and THG intensity is plotted on the log scale, the conversion efficiency of THG is an order of magnitude less than SHG. **(c)** Log$_{10}$-Log$_{10}$ plot of incident power vs SHG/THG intensity. **(d)** and **(e)** is the SHG/THG transmitted from the crystal imaged on a white screen respectively.

We tuned the input linear polarization by incorporating a half-wave plate in the excitation path. Subsequently, the recorded *x* and *y* SHG/THG intensity components, denoted as $I_x^{2\omega}, I_x^{3\omega}$ and $I_y^{2\omega} I_y^{3\omega}$, are specifically resolved in the direction of the crystal axis. This resolution is achieved by employing a linear polarizer in the collection path oriented parallel (0°) and perpendicular (90°) along the crystal axis. The polarization dependence for SHG/THG is plotted in Figure 3(a-d). The SHG pattern exhibits a six-lobe symmetry consistent with the trigonal crystal structure; however, the observed 30° angular offset between the x- and y-polarized components reveal in-plane anisotropy in the effective second-order susceptibility. THG emission patterns are anisotropic with dominant two lobe patterns and two other small lobes could be seen if the rotational resolution is increased while a minimum step size of 10° is maintained to limit data acquisition time for avoiding laser induced damage. The maximum THG emission intensity is collected at 105° and 285° along the *x*-axis and 15° and 195° along the *y*-axis. Notably, the shape of the SHG/THG emission pattern is greatly influenced by the values of the nonlinear susceptibility $\chi^2$ and $\chi^3$ tensor elements respectively. We note that the overall shapes of both SHG and THG patterns are strongly influenced by the relative magnitudes of the off-diagonal tensor components [35].

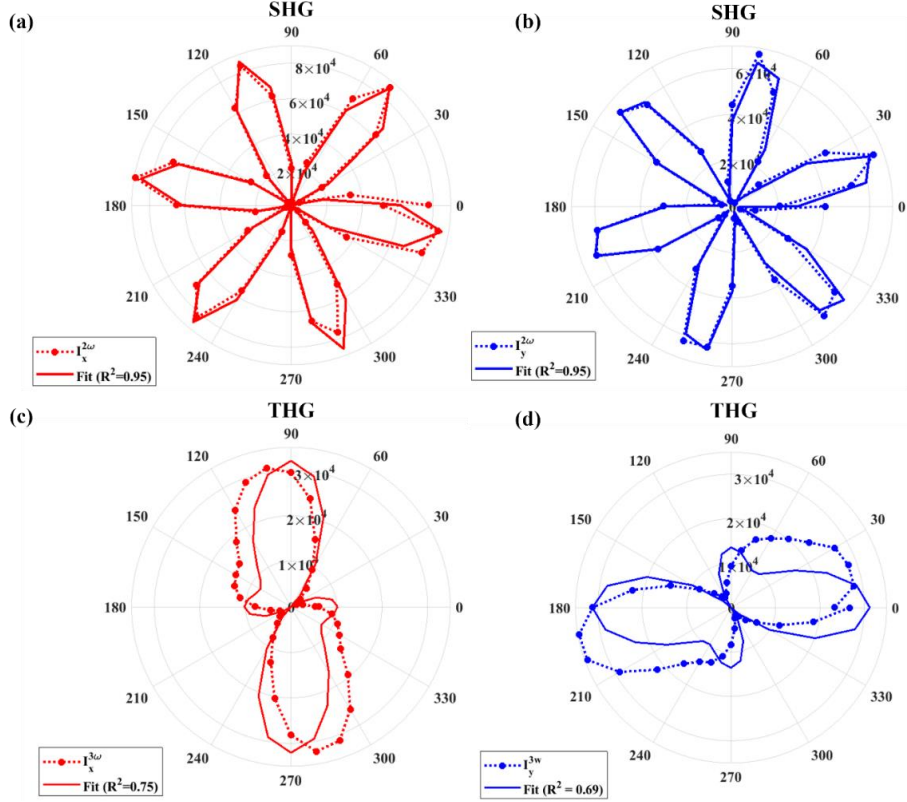

**Figure 3:** Polar plots of SH polarization dependence when the input HWP and output polarizer is rotated in parallel **(a)** and perpendicular **(b)** configurations. Solid lines represent the fits obtained from equations 10 and 11 respectively. Polar plots of TH polarization dependence when the output intensity is collected along the horizontal **(c)** and vertical **(d)** polarization states. Solid lines here also represent the fit obtained from equations 13 and 14 respectively.

To gain deeper insight into the observed SHG and THG emission patterns, we next compare the measurements with a symmetry-guided model of the second- and third-order nonlinear susceptibility. AgScP$_2$S$_6$ crystallizes in the trigonal $P3_{1c}$ space group, corresponding to the $C_3$ point-group symmetry. Under this symmetry, the $\chi^2$ tensor contains several non-vanishing elements dictated by the layered stacking geometry. These components are not independent; rather, they are constrained by the symmetry relations:

$$\chi^{(2)}_{xxx} = \chi^{(2)}_{-xyy} = \chi^{(2)}_{-yyz} = \chi^{(2)}_{-yxy}, \quad (1)$$

$$\chi^{(2)}_{xyz} = \chi^{(2)}_{-yxz}, \quad (2)$$

$$\chi^{(2)}_{xzy} = \chi^{(2)}_{-yzx}, \quad (3)$$

$$\chi^{(2)}_{xzx} = \chi^{(2)}_{yzy}, \quad (4)$$

$$\chi^{(2)}_{xxz} = \chi^{(2)}_{yyz}, \quad (5)$$

$$\chi^{(2)}_{yyy} = \chi^{(2)}_{-yxx} = \chi^{(2)}_{-xxy} = \chi^{(2)}_{-xyx}, \tag{6}$$

$$X^{(2)}_{zxx} = \chi^{(2)}_{zyy} = \chi^{(2)}_{zzz}, \text{ and} \tag{7}$$

$$\chi^{(2)}_{zxy} = \chi^{(2)}_{-zyx}, \tag{8}$$

where $x$ denotes the crystal mirror symmetry i.e., the armchair direction and $y$ is the zigzag direction in the crystal. Since the input laser excitation field is in the $x$-$y$ plane, the $\chi^2$ tensor containing $z$-component goes to zero and electric field is given as $\vec{E} = \hat{x}E\cos\theta + \hat{y}E\sin\theta$. Thus, the resultant SHG tensor is solved in similar fashion as shown in literature [36,37],

$$\begin{pmatrix} P_x^{(2\omega)} \\ P_y^{(2\omega)} \end{pmatrix} = \varepsilon \begin{pmatrix} \chi_{xxx} & -\chi_{xxx} & 0 & 0 & 0 & -\chi_{yyy} \\ -\chi_{yyy} & \chi_{yyy} & 0 & 0 & 0 & -\chi_{xxx} \end{pmatrix} \begin{pmatrix} E_x^2 \cos^2\theta \\ E_y^2 \sin^2\theta \\ 0 \\ 0 \\ 0 \\ 2E_x \cos\theta E_y \sin\theta \end{pmatrix} \tag{9}$$

By simplifying the trigonometric functions and since the linear polarizer before the detector also rotates in sync $\left(\alpha = \frac{\theta}{2}\right)$ with half-waveplate for SHG measurements. The intensity of the SHG field when the waveplate and polarizer in sync and when they are offset by 90° is given as,

$$I_x^{2\omega} \sim \{\chi_{11}[\cos(3\theta) - \sin(3\theta)] - \chi_{22}[\cos(3\theta) - \sin(3\theta)]\}^2 \tag{10}$$

$$I_y^{2\omega} \sim \left\{\chi_{11}\left[\cos\left(3\theta + \frac{\pi}{2}\right) - \sin\left(3\theta + \frac{\pi}{2}\right)\right] - \chi_{22}[\cos\left(3\theta + \frac{\pi}{2}\right) - \sin\left(3\theta + \frac{\pi}{2}\right)]\right\}^2 \tag{11}$$

The above equations are used to fit the measured SHG emission patterns as shown in Figure 3 (a) and 3 (b). It is evident that theoretical fits well with experimental data.

The magnitude of second order non-linear susceptibility in terms of optical intensity given by [38],

$$I_{2\omega} = \frac{2\omega^2 \chi_2^2 l^2}{c^3 \varepsilon_0 n_{2\omega} n_\omega^2} I_\omega^2 \left(\frac{\sin\left(\frac{1}{2}\Delta kl\right)}{\frac{1}{2}\Delta kl}\right)^2 \tag{12}$$

Where $c$ is the speed of the light in vacuum, $\varepsilon_0$ is the vacuum permittivity, $l$ is the thickness of the material, $n_{2\omega}$ and $n_\omega$ are the refractive index at second harmonic and fundamental wavelength respectively. By considering all the experimental parameters related to laser excitation and AgScP$_2$S$_6$ material mentioned in Supplementary S3, the refractive index (as we measured by ellipsometry), of the material at the fundamental wavelength $n_1$ and second harmonic wavelength $n_2$ is measured to be 1.87 and 1.81 respectively.

$\Delta k = \frac{2\pi}{\lambda_1 \lambda_2}(\lambda_1 - \lambda_2)$ represents the phase mismatch between the pump beam and the forward propagating SHG beam in transmission configuration. The total SHG emission is collected without the analyzer and using highly sensitive power meter. The measured SHG power ($P^{2\omega}$) is 31.1 µW when the input power is set to 2 mW. Thus, the $\chi^2$ value of the AgScP$_2$S$_6$ crystal is measured to be $0.82 \times 10^{-8}$ $m/V$, which is few orders of magnitude higher than recently explored vdW materials as tabulated in Table 3.

Similarly, the THG emission patterns are also fitted to a theoretical model. Since the material belongs to trigonal P31c crystal system, and their respective third order nonlinear susceptibility tensor, $\chi^3$ features non-zero elements: $\chi_{10}, \chi_{11}, \chi_{12}, \chi_{14}, \chi_{15}, \chi_{16}, \chi_{31}, \chi_{32}, \chi_{35}$. In this notation, the first subscript refers to (1=x, 2=y, 3=z) and the second subscript refers to (1=xxx, 2=yyy, 4=yzz, 5=yyz, 6=xzz, 0=xyz). Therefore, the intensity along x and y axis is given as (See Supplementary section S4 for detailed derivation),

$$I_x^{(3\omega)} \propto (\chi_{11}\cos^3\theta + \frac{1}{\chi_{11}}\cos\theta\sin^2\theta + \chi_{12}\sin^3\theta + \frac{1}{\chi_{12}}\sin\theta\cos^2\theta)^2 \quad (13)$$

$$I_y^{(3\omega)} \propto (\chi_{11}\sin^3\theta + \frac{1}{\chi_{11}}\sin\theta\cos^2\theta - \chi_{12}\cos^3\theta - \frac{1}{\chi_{12}}\cos\theta\sin^2\theta)^2 \quad (14)$$

We employed equations 13 and 14 to fit the measured THG emission patterns, and the theoretical fits are depicted with solid lines in the corresponding colors in Figure 3 (c) and (d). It is evident that the theoretical fits well with the experimental data. Also, the relative magnitudes of $\chi_{11}$ & $\chi_{12}$, are obtained by the theoretical fit, with their average values measured to be $\chi_{11}:\chi_{12} = 1:0.66$.

The third-order nonlinear susceptibility of AgScP$_2$S$_6$ crystal can be further extracted by the following expression [33],

$$|\chi^3|^2 = \left[ \frac{16(n_3^2 + k_3^2)^{1/2} n_1^3 \varepsilon_o^2 c^4 f_{rep}^2 W^4 \tau^2 \left[\frac{\pi}{4ln2}\right]^3 P^{3\omega}}{9\omega^2 l^2 (P^{3\omega})^3} \left( \frac{\left(\frac{4\pi^2 k_3^2 l^2}{\lambda_3^2} + \Delta k^2 l^2\right)}{e^{-\frac{4\pi k_3 l}{\lambda_3}} - 2\cos(\Delta kl) e^{-\frac{2\pi k_3 l}{\lambda_3}} + 1} \right) e^{\frac{4\pi k_3 l}{\lambda_3}} \right] \quad (15)$$

By considering all the experimental parameters same as $\chi^2$ and the refractive index of the third harmonic signal is measured to be n=3.86 from section S3. The measured THG power ($P^{3\omega}$) is 10.5 nW when the input power is set to 2 mW. Thus, the $\chi^{(3)}$ value of the AgScP$_2$S$_6$ crystal is measured to be $1.69 \times 10^{-17} m^2/V^2$, which is few orders of magnitude higher than recently explored vdW materials. Moreover, the estimated $\chi^3$ value for the AgScP$_2$S$_6$ crystal derived from equation 15 corresponds to the $\chi_{11}$ element, as it is based on the measured THG emission power with the incident linear polarization along the x-axis. From this the relative magnitude of $\chi_{12}$ is given as $1.12 \times 10^{-17} m^2/V^2$.

Table 3: Comparison of 2nd/3rd order non-linear susceptibility with respect to other novel materials.

| Material | Wavelength (nm) | Thickness | $\chi^2$ (m/V) | $\chi^3$ (m$^2$/V$^2$) | Reference |
|---|---|---|---|---|---|
| Graphene | 1560 | ~ 1 layer |  | $4 \times 10^{-15}$ – $1.5 \times 10^{-19}$ | [36] [40] |
| MoS$_2$ | 1560 | ~ 1 layer | $2.9 \times 10^{-11}$ | $2.4 \times 10^{-19}$ | [40,41] |
| MoSe$_2$ | 1620/1560 | ~ 1 layer | $5.0 \times 10^{-11}$ | $2.2 \times 10^{-19}$ | [39,42] |
| WS$_2$ | 832/1560 | ~ 1 layer | $4.5 \times 10^{-9}$ | $2.4 \times 10^{-19}$ | [41,43] |
| GaTe | 1560 | Multilayer | $1.15 \times 10^{-12}$ | $2.0 \times 10^{-16}$ | [44] |
| PdPSe | 1300 | 6 layers | $6.4 \times 10^{-11}$ | $6.2 \times 10^{-19}$ | [45] |
| BP | 1560 | Multilayer |  | $1.6 \times 10^{-19}$ | [46] |
| SnSe$_2$ | 1560 | Multilayer |  | $4.1 \times 10^{-19}$ | [47] |
| NbOI$_2$ | 1560 | 22 nm |  | $0.9 \times 10^{-18}$ | [48] |
| RPP (Br$_{n=1}$) | 1210 | 22 nm |  | $4.7 \times 10^{-18}$ | [49] |
| LiN | 1064 | Bulk | $1.5 \times 10^{-11}$ | $3.9 \times 10^{-18}$ | [50,51] |
| AgScP$_2$S$_6$ | 1030 | 9 μm | $0.82 \times 10^{-8}$ | $1.7 \times 10^{-17}$ | This work |

## 2.3 Temperature Dependence

Temperature-dependent SHG and THG were performed in AgScP$_2$S$_6$ utilizing a cryogenic chamber in vacuum. The laser parameters and collection methods were kept the same as mentioned before (see supplementary S6 (d) section for details), and the harmonic intensity is along $x$ and $y$ components of the polarization are recorded when the material is cooled from 300 K (room temperature) to 25 K in transmission setup to minimize the surface contributions. From Figure S4 & S5 the emission intensity increases exponentially with temperature indicating in-plane structural electronic changes. Polarization input was controlled with a half-waveplate and polarizer set up to look at parallel polar input to output and orthogonal polar input to output. This allowed us to effectively rotate along the stacking axis of the crystal by rotating the input polarization using the half-waveplate and select parallel and perpendicular outputs using the analyzer after the crystal. Strong SHG emission across the entire temperature range in bulk AgScP$_2$S$_6$ supports our structural refinements given SHG is forbidden to inversion centric space groups which literature assigns to this system. SHG/THG emission drops drastically with reducing temperature. This behavior, which may be due to cooling, can lead to reduced carrier mobility, i.e., if the material's carrier density decreases at lower temperatures, the efficiency of non- linear processes, including harmonic signals will be affected [52]. Notably no change in the symmetry of the SHG/THG polar plot is present, again supporting the conclusion that there is no low temperature phase transition in this material. A slight angular drift in the lobe locations is seen, but this is believed to be due to experimental errors. A maximal drift of 15° is seen. Crossed and parallel data for SHG are rotated relative to each other by 30°, which would seem to be representative of the cation arrangement in the system. Other MTPs such as AgInP$_2$S$_6$ and antiferromagnetic compounds of MPS$_3$ (M=Ni, Fe, and Mn) show similar 6-lobed SHG polar plots. Crossed and parallel data for THG are rotated relative to each other by ~90°.

Next, the anisotropy ratio of the material is examined using SHG/THG emission which is used to infer how much the materials property varies along different crystallographic axes. The anisotropy ratio is calculated with change in temperature which measures the SHG/THG intensity when excitation polarization is along the two orthogonal directions. The SHG anisotropy ratio is related to the $\chi^2$ tensor which is the ratio of intensity measured along $\chi_{11}$ and $\chi_{22}$ direction. From equations 10 and 11 it can be directly calculated using the fitting results. Similarly, for THG anisotropy ratio is related to the tensor elements $\chi_{11}$ and $\chi_{12}$ which is also calculated directly from the fit results using equations 13 and 14. From the Figure 4 Temperature-dependent polarization-resolved measurements reveal that the third-harmonic generation (THG) anisotropy ratio remains nearly constant at approximately 0.7 across the entire temperature range, indicating a moderately anisotropic yet thermally robust third-order nonlinear response. In contrast, the second-harmonic generation (SHG) exhibits a pronounced anisotropy, with an anisotropy ratio of ~0.1 at most temperatures, reflecting a highly directional second-order susceptibility. Interestingly, a marked increase in the SHG anisotropy ratio to ~0.7 is observed between 200 K and 250 K, suggesting a significant reduction in anisotropy. As no structural phase transitions are detected via Raman spectroscopy or differential scanning calorimetry (DSC), this change is likely due to subtle temperature-induced modifications in the electronic structure or lattice dynamics that affect the balance of tensor components contributing to the SHG response. Similar phenomena have been observed in other studies where strain engineering in two-dimensional materials led to significant alterations in SHG behavior without apparent phase transitions. The uniaxial strain developed in the material could break in-plane symmetry, leading to anisotropic SHG patterns in monolayer materials [53]. Additionally, the anisotropic electron and lattice dynamics can influence excitonic properties and coherent phonon generation, impacting the nonlinear optical responses in materials like Ta$_2$NiSe$_5$ [54]. Thus, the sensitivity of SHG to subtle changes in lattice dynamics and electronic structure, supporting the interpretation that the observed temperature-dependent anisotropy alterations are due to such effects rather than macroscopic phase transitions.

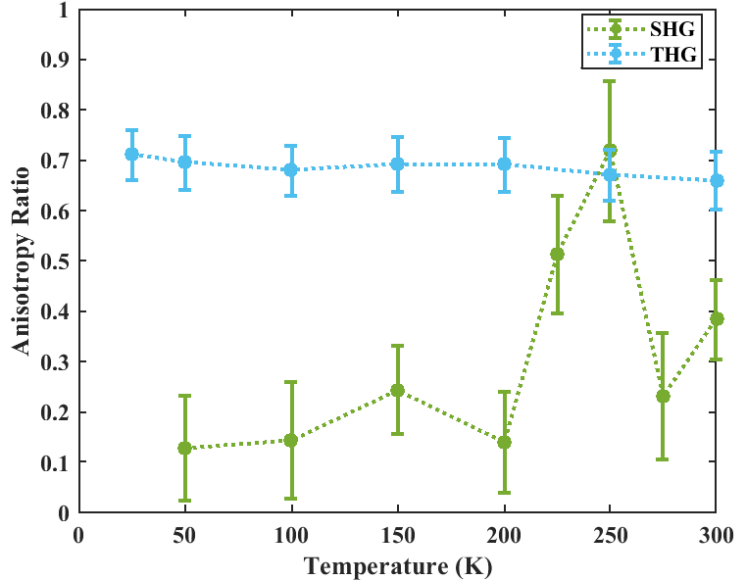

**Figure 4:** Anisotropy ratio of SH along $\left(\frac{\chi_{22}}{\chi_{11}}\right)$ and TH along $\left(\frac{\chi_{12}}{\chi_{11}}\right)$ tensor elements.

### 2.4 Polarization State Analysis

The analysis of the polarization state of SH/TH emission from 2D materials provides crucial insights into their NLO properties. When a fundamental laser beam interacts with a 2D material, the polarization state of the resulting harmonic emission can reveal information about the material's symmetry and the orientation of its nonlinear susceptibility tensor elements. In this section, the polarization ellipticity and polarization orientation of SH/TH emission with respect to the pump polarization orientation for the ~9 μm thick AgScP$_2$S$_6$ crystal is analyzed by measuring the Stokes parameters of SH and TH emission. The Stokes parameters $S_0$, $S_1$, $S_2$, and $S_3$ are measured by passing the SH/TH emission through a series of optical elements, including polarizers and wave plates, and recording the intensity of the light. These parameters describe the total intensity ($S_0$), the degree of linear polarization ($S_1$ and $S_2$), and the degree of circular polarization ($S_3$). Then the recorded intensity data is fitted to a function of the form given below from which the Stokes parameters are obtained.

$$I(\theta) = S_0 + S_1 \cos(4\theta) + S_2(\sin 4\theta) + S_3(\sin 2\theta) \tag{16}$$

Then the polarization angle (or orientation of the major axis of the polarization ellipse) is given by:

$$\theta_p = \frac{1}{2}\tan^{-1}\left(\frac{S_2}{S_1}\right) \tag{17}$$

and the ellipticity angle which represents the shape of the polarization ellipse is defined as:

$$\varepsilon = \frac{1}{2}\sin^{-1}\left(\frac{S_3}{S_0}\right) \qquad (18)$$

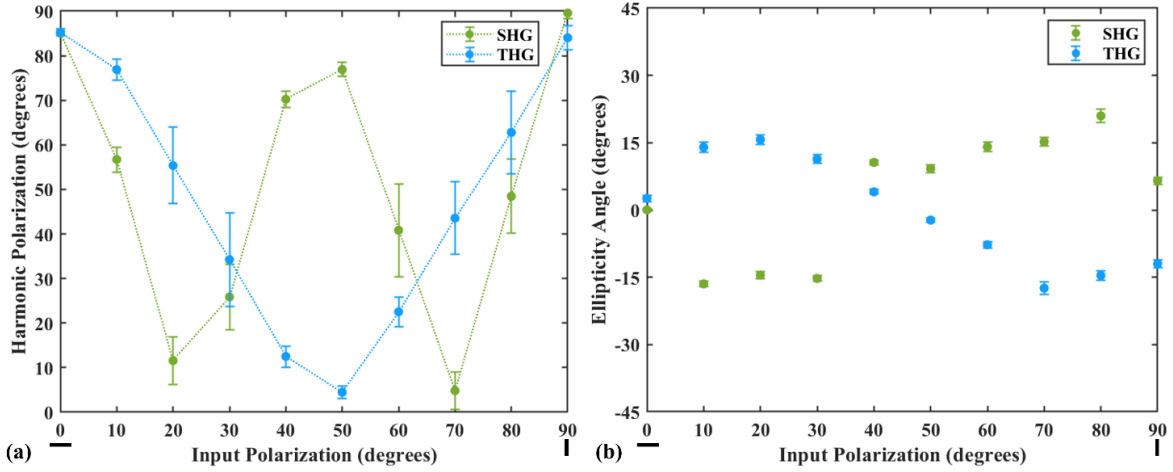

**Figure 5. (a), (b)** Dependence of SH/TH polarization orientation and SH/TH ellipticity with respect to input polarization respectively.

The relationship between the polarization orientation of the SH/TH emission and the rotation of the pump polarization from 0° to 90° is illustrated in Figure 5 (a). The SH polarization orientation is represented in green circles, and the TH polarization orientation is represented in blue circles. The change in harmonic polarization orientation with respect to input polarization in both harmonic signals is likely a direct consequence of the intrinsic optical anisotropy of the material. Since 2D materials often exhibit strong in-plane optical anisotropy, the nonlinear polarization response is expected to be highly sensitive to the input polarization direction. We can infer form the figure that both the SH and TH are in vertical polarization state to start with when the input polarization is in horizontal direction which means the harmonic generation is due to Type I phase matching conditions. The SH emission switches close to horizontal polarization when the input is at 20 and 70 degrees, whereas the TH emission switches to horizontal polarization state when the input polarization angle is at ~50 degrees.

Additionally, Figure 5 (b) shows the ellipticity angle of the SH/TH emission as a function of the pump beam's polarization orientation. The ellipticity angle describes how much the polarization deviates from being purely linear, with zero indicating perfectly linear polarization and increasing values corresponding to elliptical polarization states. Both SH and TH emissions produce a decent amount of ellipticity suggesting that birefringence effects and phase retardations are present within the material, possibly due to local strain or multilayer interference effects in the material.

The polarization revolved spectral intensity maps of SH and TH intensity as functions of wavelength and input half-wave plate angle are shown in Figure S6. These figures give a comprehensive view of how harmonic generation efficiency changes with polarization tuning from horizontal to vertical. In Figure S6 (a) and (b) though the individual polarization directions of SHG shows significant modulation with respect to the input HWP angle, the combined SHG intensity remains approximately constant indicating that the process predominantly redistributes the harmonic signal between polarization states without substantially altering the total SHG efficiency. Similarly, from figure S6 (c) and (d) THG also follows the same trend as SHG where the THG efficiency remains approximately constant with change in input polarization.

Next, the evolution of SH and TH intensity and ellipticity under elliptically polarized pump beam is investigated. The elliptical polarization of the pump beam is controlled by a rotating quarter-wave plate (QWP) positioned before the sample, with the incident linear polarization set horizontal to the material. The ellipticity $\epsilon$ of the input pump beam is gradually adjusted from 0° (horizontal polarization) to $\pm 45°$ (right circular polarization, RCP, left circular polarization, LCP) by rotating the fast axis of the QWP ($\theta$) with respect to the $x$-axis. Figure 6 displays the harmonic ellipticity as a function of the input QWP angle, revealing a strong dependence of both SHG and THG on the input polarization ellipticity. The SHG ellipticity follows the input ellipticity direction, increasing monotonically from negative to positive as the QWP angle varies from –45° to +45°. In contrast, the THG ellipticity shows an inverted trend, decreasing with an increasing QWP angle. This opposing behavior highlights the distinct polarization selection rules and nonlinear tensor contributions involved in SHG and THG processes.

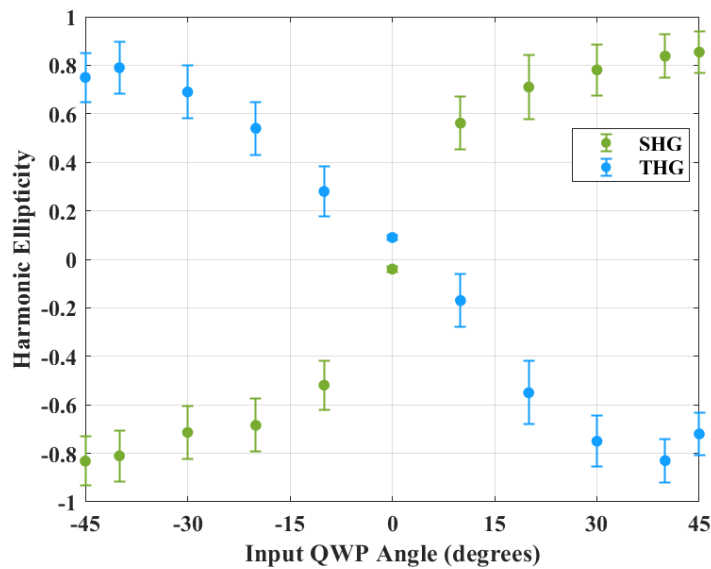

**Figure 6:** Dependence of SH/TH ellipticity with respect to input QWP angle.

Figure 7 shows the spectrally resolved SHG and THG intensities as a function of the input QWP angle, which effectively tunes the input polarization from linear to circular. From figure 7 (a) the SHG signal exhibits a pronounced intensity enhancement around QWP angles of 45° and 135°, which correspond to right- and left-handed circular polarization, as indicated by the circular markers. In contrast, minimum in SHG intensity occurs at QWP angles of 0°, 90°, and 180°, where the input polarization is linear. This behavior suggests that the SHG process in this material is strongly enhanced under circularly polarized input. Such dependence is consistent with SHG in systems where the nonlinear susceptibility tensor elements couple efficiently to circular polarization, or where spin orbit like interactions in 2D or non-centrosymmetric materials enhance nonlinear conversion for specific helicities [55]. Figure 7 (b) shows the THG response under the same QWP variation. Here, the THG intensity peaks at QWP angles of 0°, 90°, and 180° i.e., under linearly polarized input and drops significantly near 45° and 135°, where the input becomes circularly polarized. This inverse trend compared to SHG indicates that the THG process is more efficient when the driving field is linearly polarized. This is expected in many systems where THG originates from third-order tensor components ($\chi^3$) that are dominant for linearly polarized excitation due to the preserved field alignment across polarization cycles.

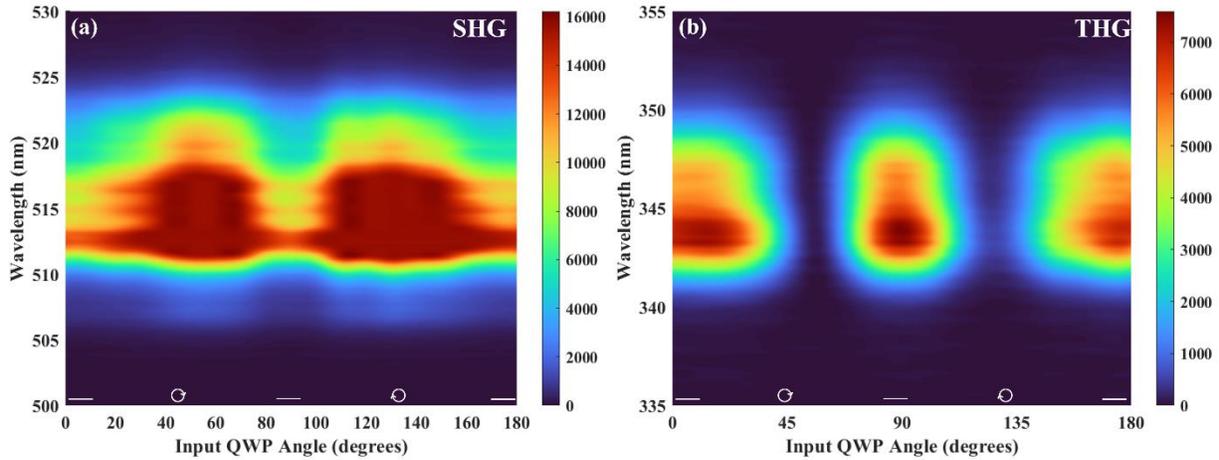

**Figure 7: (a), (b)** are the SH/TH intensity when the input QWP is rotated from 0 to 180 degrees.

3. Conclusions

AgScP$_2$S$_6$ emerges from this study as a structurally robust and highly efficient nonlinear optical material within the family of van der Waals layered crystals. SHG measurements verify its non-centrosymmetric P3$_{1c}$ symmetry with racemic twinning, while temperature-dependent behavior shows that Ag/Sc cation ordering remains stable down to low temperatures. This structural resilience differentiates AgScP$_2$S$_6$ from many anisotropic 2D materials that suffer from ambient instability or symmetry degradation. The material exhibits extraordinarily large second- and third-order nonlinear responses, with SHG and THG efficiencies far exceeding those of most reported vdW materials when normalized to thickness. Polarization-resolved harmonic measurements confirm the presence of strong in-plane anisotropy consistent with $C_3$ symmetry and highlight the role of off-diagonal $\chi^2$ and $\chi^3$ tensor elements in shaping the SHG/THG emission profiles. Additionally, the marked temperature tunability of the harmonic yield suggests that coupling between lattice dynamics and electronic structure can be leveraged to modulate nonlinear conversion efficiency. Our results establish that $AgScP_2S_6$ combines ambient stability, high optical anisotropy, and exceptionally strong harmonic generation, positioning it as a leading candidate for emerging nonlinear photonic technologies. Beyond advancing the understanding of nonlinear processes in metal thiophosphates, the findings open pathways toward practical implementation. Future work should investigate thickness-dependent nonlinearities, explore excitonic or phonon-coupled contributions to harmonic generation, and integrate this material into on-chip architectures for polarization-programmable or thermally reconfigurable frequency converters. Such efforts will further unlock the potential of $AgScP_2S_6$ for next-generation nanoscale photonic devices.

**Acknowledgements**

The authors acknowledge support through the United States Air Force Office of Scientific Research (AFOSR) LRIR23RXCOR003, LRIR26RXCOR010, FA9550-25-1-0297 and AOARD-NSTC Grant Number F4GGA21207H002. The work at the Ohio State University is also partially supported by the President's Research Excellence (PRE) Program: Catalyst funds. The work at University of Texas at Dallas is supported by US Air Force Office of Scientific Research Grant No. FA9550-19-1-0037, National Science Foundation- DMREF-2324033, and Office of Naval Research grant no. N00014-23-1-2020.

**Disclosures**

The authors declare no conflicts of interest.

**Data Availability**

Data underlying the results presented in this paper are not publicly available at this time but may be obtained from the authors upon reasonable request.

**Supplemental Information**

See Supplemental document for supporting content.